\documentclass[reprint,graphicx,superscriptaddress,nofootinbib]{revtex4-1}
\usepackage{epsfig}
\usepackage{color}
\usepackage{amsmath}
\usepackage{amssymb}
\usepackage{mathtools}
\usepackage{longtable}
\usepackage{dcolumn}% Align table columns on decimal point
\usepackage{bm}% bold math
\usepackage{float}
\usepackage{multirow}% http://ctan.org/pkg/multirow
\usepackage{hhline}% http://ctan.org/pkg/hhline
\usepackage{lipsum}

\providecommand{\abs}[1]{\lvert#1\rvert}

\newcommand{\Expect}{{\rm I\kern-.3em E}}
\DeclarePairedDelimiter{\ceil}{\lceil}{\rceil}

\draft % marks overfull lines with a black rule on the right

\begin{document}

%\preprint{}

%\title{Programming and tuning a quantum annealing device to solve real-world applications} %Title of paper
\title{A Performance Estimator for Quantum Annealers: \\Gauge selection and Parameter Setting. } %Title of paper
%\title{Gauge selection and (Embedding) Parameter Setting for Quantum Annealing Devices: A Performance Estimator Approach } %Title of paper
%\title{A Performance Estimator for Gauge Selection and Parameter Setting in Quantum Annealers.} %Title of paper
%\title{A Performance Estimator Approach for Gauge Selection and Parameter Setting in Quantum Annealing Devices.} %Title of paper



\author{Alejandro Perdomo-Ortiz}
\email[Corresponding author's e-mail: ]{alejandro.perdomoortiz@nasa.gov}
%\thanks{}
\affiliation{Quantum Artificial Intelligence Lab., NASA Ames Research Center, Moffett Field, CA 94035, USA}
\affiliation{University of California Santa Cruz at
NASA Ames Research Center, Moffett Field, CA 94035, USA}


\author{Joseph Fluegemann}
%\thanks{}
\affiliation{Quantum Artificial Intelligence Lab., NASA Ames Research Center, Moffett Field, CA 94035, USA}
\affiliation{San Jose State Research Foundation at
NASA Ames Research Center, Moffett Field, CA 94035, USA}


\author{Rupak Biswas}
%\thanks{}
\affiliation{Exploration Technology Directorate, NASA Ames Research Center, Moffett Field, CA 94035}


\author{Vadim N. Smelyanskiy}
%\thanks{}
\affiliation{Quantum Artificial Intelligence Lab., NASA Ames Research Center, Moffett Field, CA 94035, USA}


\date{\today}

\begin{abstract}
With the advent of large-scale quantum annealing devices, several challenges have emerged. For example, it has been shown that the performance of a device can be significantly affected by several degrees of freedom when programming the device; a common example being gauge selection. To date, no experimentally-tested strategy exists to select the best programming specifications. We developed a score function that can be calculated from a number of readouts much smaller than the number of readouts required to find the desired solution. We show how this performance estimator can be used to guide, for example, the selection of the optimal gauges out of a pool of random gauge candidates and how to select the values of parameters for which we have no \textit{a priori} knowledge of the optimal value. For the latter, we illustrate the concept by applying the score function to set the strength of the parameter intended to enforce the embedding of the logical graph into the hardware architecture, a challenge frequently encountered in the implementation of real-world problem instances. Since the harder the problem instances, the more useful the strategies proposed in this work are, we expect the programming strategies proposed to significantly reduce the time of future benchmark studies and in help finding the solution of hard-to-solve real-world applications implemented in the next generation of quantum annealing devices.
\end{abstract}

\maketitle %\maketitle must follow title, authors, abstract and \pacs

\section{Introduction}

The fabrication of scalable hardware architectures for quantum annealers~\cite{Bunyk_IEEE2014,johnson_quantum_2011} to solve discrete optimization problems has sparked interest in quantum annealing algorithms~\cite{kadowaki_quantum_1998,Farhi2001}. Current research studies focus on both fundamental and practical important questions, including the implementation of real-world applications~\cite{PerdomoOrtiz2012_LPF, Gaitan2012, PerdomoOrtiz_EPJST2015, RieffelQIP2015, OGormanEPJST2015}, defining criteria for detecting quantum speedup and the computational role of quantum tunneling~\cite{Ronnow25072014,Boixo_arXiv2015}, proposals for error-supression schemes~\cite{Pudenz_NatCom2014},  benchmark studies comparing classical and quantum annealing~\cite{boixo_NatCommun2013, boixo_evidence_2014,Shin_arXiv2014,  Albash_EPJST2015, MartinMayor_arXiv2015, Hen_arXiv2015,King_arXiv2015}, and using spin-glass perspectives into the hardness of computational problems studied~\cite{Katzgraber_PRX2015,Venturelli_arXiv2014}.

The next generation of quantum annealers will likely allow for the exploration of harder and more interesting problems instances. Even in the case of a quantum processor with only $\sim$500 qubits, one could already see the appearance of some hard to solve instances, for which the optimal solution was not found out of a few thousand annealing cycles~\cite{Ronnow25072014}. It is precisely for these hard instances that the methods developed in this paper are the most useful, since they allow us to extract the best programming settings enhancing the probability of finding the ground state, therefore reducing significantly the time to solution for both future benchmark studies and for real-world applications.

The first step of solving a problem using a quantum annealer is to map the problem to the hardware architecture. The quantum hardware employed consists of 64 units of a recently characterized eight-qubit unit cell~\cite{harris2010,johnson_quantum_2011}. Post-fabrication characterization determined that only 509 qubits out of the 512 qubit array can be reliably used for computation (Fig.~\ref{fig:chimera} in Appendix~\ref{sec:progQA}). The array of coupled superconducting flux qubits is, effectively, an artificial Ising spin system with programmable spin-spin couplings and transverse magnetic fields. It is designed to solve instances of the following (NP-hard~\cite{Barahona1982}) classical optimization problem: Given a set of local longitudinal $\{h_i\}$ and an interaction matrix $\{J_{ij}\}$, find the assignment $\mathbf{s^*} = s^*_1 s^*_2 \cdots s^*_N$, that minimizes the objective function $E(\mathbf{s})$, 
\begin{equation}\label{eq:Eising}
E_{\rm{ising}}(\mathbf{s})  = \sum_{1 \le i \le N} h_{i} s_i  + \sum_{1 \le i<j\le N} J_{ij} s_{i} s_{j},
\end{equation}
where, $\abs{h_i} \le 2$, $\abs{J_{ij}} \le 1$, and $s_i \in \{+1,-1\}$. Finding the optimal $\mathbf{s^*}$ is equivalent to finding the ground state of the corresponding Ising Hamiltonian, $H_{p}  =  \sum_{i} h_{i}\sigma_{i}^{z}  + \sum_{i<j} J_{ij}\sigma_{i}^{z} \sigma_{j}^{z}$
where $\sigma_{i}^{z}$ are Pauli matrices acting on the $i$th spin.

Physical realizations of quantum annealing come with certain degrees of freedom affecting the performance of the quantum annealing device. Each realization of such degrees of freedom determine a unique \textit{Hamiltonian specification} or realization of Eq.~\ref{eq:Eising}. Although there is some understanding of the several factors affecting the performance of quantum annealing devices \cite{King_arXiv2014}, there is a need for concrete scalable strategies coping with the analog control error (ACE) intrinsic to physical hardware implementations. For example, to the best of our knowledge there is no known``rule-of-thumb" in the selection of such parameters,  thus motivating our study. (we rule out the only one wide spread in the community, in Appendix~\ref{app:thumbrule}.)

In the absence of noise or any miscalibration, the performance of the quantum device should be the same under any gauge realization~\cite{boixo_evidence_2014}. Previous studies show that the current generation of D-Wave devices with hundreds of qubits is very sensitive to this selection~\cite{boixo_evidence_2014,PerdomoOrtiz_EPJST2015}. Some other degrees of freedom correspond to parameters we do not yet know \textit{a priori} how to set. This is the case for penalty strength in the construction of quadratic unconstrained binary optimization (QUBO) Hamiltonians \cite{PerdomoOrtiz_EPJST2015,Babbush2014} or penalties associated with the strength of the set of qubits defining a logical qubit in the QUBO graph to hardware graph procedure~\cite{Cai-14}.

In Sec.~\ref{sec:tuningQA} we present a strategy for tuning and optimizing a quantum annealing algorithm, finding the best parameters out of a pool of candidates and selecting the Hamiltonian specifications with the best performance. It is in this section where most of the new results are presented. For accessibility to the readers, we divided this section into two main threads. Readers interested only in gauge selection (such as those researchers interested in benchmark studies, for example, on random spin-glass instances) can find the procedures needed in Sec.~\ref{subsec:pielite} - \ref{subsec:gaugeselect}. For readers interested in more general real-world application instances, where parameters for embedding procedures and other penalties need to be set, we devote Sec.~\ref{subsec:tuneJE} to discussing the adjustments to the technique to deal with these additional challenges. In Sec.~\ref{sec:conclusions}, we delineate some future directions and possible further applications of the present work.

\section{Tuning a Quantum Annealing Algorithm}\label{sec:tuningQA}

As previously discussed, there are many degrees of freedom at the time of programming a quantum annealing device to solve a specific problem instance. Each realization of such degrees of freedom determine what we call a \textit{Hamiltonian specification} for the quantum annealing cycles. For the purpose of generality, we leave the discussion at a very high-level form and in the following sections we will present application examples in different common practical scenarios, e.g., gauge selection and setting the strength of couplings among physical qubits representing a qubit from the original logical graph, also known as the embedding parameter setting problem~\cite{Choi2008}. In this general framework presented here, we only need to keep in mind that the performance of the device is determined by the programming degrees of freedom through the different Hamiltonian specifications. The main question we discuss next: How do we select the Hamiltonian realization that yields the best performance of the device? It is the focus of this section to answer this question with a procedure requiring a minimum overhead, as described next. 

\subsection{Performance Estimator: The Elite Mean, $\Pi_{\rm{elite}}$}\label{subsec:pielite}

Assume that for each Hamiltonian specification you can easily request a total number of readouts $N_{\rm{reads}}$ from the quantum annealer. In some cases, these $N_{\rm{reads}}$ must be obtained in batches due to programming limitations. For example, in the current D-Wave Two processor hosted at NASA Ames, programming the device with an annealing time per cycle of $t_a = 100 \mu s$ allows for a maximum number of readouts of 10,000. If the device is operated at $t_a = 20 \mu s$ this maximum number is 50,000~\footnote{In our machine, there is a maximum duty time per submission. This value is set to $10^6 \mu s$, the reason why the maximum number of readouts at $t_a = 20 \mu s$ is $50,000$ while only $10,000$ at $t_a = 100 \mu s$}. Therefore, while at $t_a = 20 \mu s$ a goal of $N_{\rm{reads}} = 50,000$ can be obtained in one shot, at $t_a = 100 \mu s$ we need to request 5 repetitions of 10,000 each. Let's denote the number of repetitions needed by $n_{\rm{reps}}$ and therefore the number of readouts in each repetition is $n_{\rm{reads}} = N_{\rm{reads}}/n_{\rm{reps}}$.

For each readout, $\mathbf{s}^{(i)}$ there is a corresponding $E_{\rm{ising}}(\mathbf{s}^{(i)})$ (Eq.~\ref{eq:Eising}). Let's define by $\mathbf{\tilde{E}_{\rm{ising}}}$ as the array containing the $n_{\rm{reads}}$ sorted energies, i.e.,  $\mathbf{\tilde{E}_{\rm{ising}}} = \{e_1, e_2, \cdots, e_{n_{\rm{reads}}}\}$ such that $e_i \le e_j$ for all $j > i$.  Define $\pi^{\epsilon \%}_{\rm{elite}}$ as the negative of the mean value of the lowest $\epsilon$ percent of the energies in $\mathbf{\tilde{E}_{\rm{ising}}}$. Since the array $\mathbf{\tilde{E}_{\rm{ising}}}$ contains $n_{\rm{reads}}$ sorted energies from lowest to highest, then this expectation value is equivalent to calculating the mean value using the first $n_{\rm{elite}} = \ceil{\epsilon * n_{\rm{reads}}/100}$ values in $\mathbf{\tilde{E}_{\rm{ising}}}$. Formally defined,
\begin{equation}\label{eq:pielite}
\pi^{\epsilon \%}_{\rm{elite}}(n_{\rm{reads}})  = -\sum^{n_{\rm{elite}}}_{i=1} e_i
\end{equation}   

 Since only a fixed percent of the lowest energy values are included in the calculation, we refer to this score function hereafter as the \textit{elite mean}. This expression can be generalized to the case where several repetitions are used to collect the desirable total number of samples $N_{\rm{reads}}$ by defining
\begin{equation}\label{Pielite}
\Pi^{\epsilon \%}_{\rm{elite}}(N_{\rm{reads}},n_{\rm{reps}}) = \frac{1}{n_{\rm{reps}} } \sum_{i=1}^{n_{\rm{reps}}} \pi^{\epsilon \%}_{\rm{elite}}\left(\frac{N_{\rm{reads}}}{n_{\rm{reps}}},i\right). 
\end{equation}

The minus sign in the definition Eq.~\ref{eq:pielite} gives $\pi^{\epsilon \%}_{\rm{elite}}$ the interpretation of a score function or a performance estimator; the higher its value, the better the expected performance.
Suppose one has several quantum annealers or several Hamiltonian specifications to choose from. If one is interested in assessing the performance of the device with a number of reads $N_{\rm{reads}} \ll R_{.99}$ (where $R_{.99}$ is defined as the number of readouts needed to find the desired solution at least once with a 99\% probability), we will show that $\Pi^{\epsilon \%}_{\rm{elite}}$ serves as an effective score function or performance estimator that can be used to rank and to select the best of available quantum annealing specifications to solve the problem at hand. 
The intuition for this score function follows from what is expected of a quantum annealing device: when given a problem to be solved, the quantum annealers (or the Hamiltonian specifications) that give the lower energy solutions are preferable, since a quantum annealer is designed to sample from the lowest energy configurations. Therefore, the quantum annealer specification with the lower \textit{elite mean energy} (or higher elite-mean score $\Pi^{\epsilon\%}_{\rm{elite}}$), will give better performance.

\subsection{Performance Rank}\label{subsec:ranking}

In quantum annealing, the most natural gold standard for assessing performance is the probability of observing the ground state, since it translates into the probability of finding the optimal solution to the optimization problem studied. In more precise terms, let's define the success probability of our quantum annealing algorithms by $p_s = n_{\rm{gs}}/N_{\rm{total}}$, where $n_{\rm{gs}}$ corresponds to the number of observed ground states in the total number of requested readouts $N_{\rm{total}}$. Given $p_s$, the number of repetitions needed to observe the optimal solution at least once with a 99\% probability, $R_{.99}$ is given by \cite{boixo_evidence_2014},
\begin{equation}
R_{.99} = \ceil*{\frac{\ln (1-0.99)}{\ln (1-p_s)}}
\end{equation}

The instances that will benefit the most from our selection approach are those hard-to-solve instances with a very low probability of obtaining the ground state, say with an $R_{.99}$ in the order of hundreds of millions or hundreds of millions like the example discussed in Sec.~\ref{subsec:tuneJE} ($R_{.99} \gg N_{\rm{reads}}$). The purpose of this work is to show a correlation of the performance estimator and the real performance of the machine. But how do we define or assess real performance when the number of ground states is not reasonably attainable for all the Hamiltonian configurations we explored? Take for example the instance in Sec.~\ref{subsec:tuneJE}. The default setting of the device does not provide even a single ground state solution after $N_{\rm{total}} = 50 \times 10^6$ readouts! To calculate $p_s$ for all gauges we would need to run for all 100 gauges being considered at least $N_{\rm{total}} > 50 \times 10^6$, which is beyond the scope of this work. In this work we explore two definitions of performance that allow us to rank different Hamiltonian configurations even in the case where the ground state is not obtained after a significantly large $N_{\rm{total}}$. The first and natural criteria is a \textit{greedy-like performance rank}. This method gives a lower (better) rank to a Hamiltonian specification with a lower energy. In the common case of ties, they are broken by looking at the frequency (number of occurrences) of their lowest energy state. In case these are the same, the next lowest energy is compared and if they are still the same, one compares their frequencies. The process continue until ties are broken, providing winners that are accordingly ranked lower. This method allows us to assign a unique ranking to any Hamiltonian specification whether or not we measured any ground states. Notice that in the particular case where the ground state is obtained for all the Hamiltonian configurations explored, the performance rank will still assign lower ranks to Hamiltonian specifications with larger values of $p_s$, as desired, while breaking any ties that exist.

\subsection{Gauge selection: Case study with a random spin-glass instance}\label{subsec:gaugeselect}

\begin{figure*}
\centering
\includegraphics[width=0.98\textwidth]{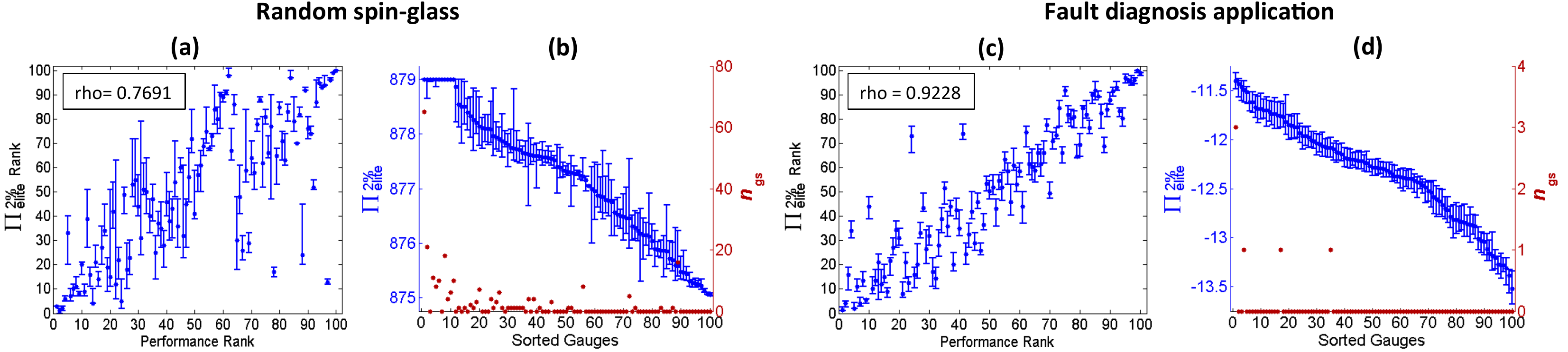}
\caption{\scriptsize{\textbf{Correlation between $\Pi_{\rm{elite}}$ rank vs. performance rank:}. (a-d) We refer to ``one experiment" (a single experiment) as the ranking of the 100 gauges according to the values of $\Pi^{2\%}_{\rm{elite}}$. In each of the 40 (100) experiments for the RS (50M-DMF) problem instance, each of the 100 gauges is scored and ranked according to $\Pi^{2\%}_{\rm{elite}}$, obtained from $N_{\rm{reads}}=50,000$ per gauge. The median value of these experiments is shown as the middle point and the limits of the error bars correspond to the 25- and 75-percentile from these experiments at $t_a  = 20 \mu$s. The number of readouts ($N_{\rm{reads}}=50,000$) used to calculate $\Pi^{2\%}_{\rm{elite}}$ rank in both examples is at least 100 times less than the number of repetitions required to find at least one solution with a probability of 99\%, (i.e., $R_{.99} \gg N_{\rm{reads}}$). (a,c) The Spearman coefficient (rho) shows a moderately strong correlation of our estimator rank and the gold-standard rank used to define the performance.  (b,d) Shown are the total number of ground states, $n_{\rm{gs}}$, across all extensive runs from $E_{\rm{ising}}$ ($E_{\rm{QUBO}}$ and using majority voting), with $N_{\rm{total}}=2$ millions ($N_{\rm{total}} = 5$ million) for the RS (50M-DMF) problem instance. It is worth mentioning that the case of no-gauge with a performance rank of 77 out of 100 in panel (c) not even a single ground state was obtained after 50 millions readouts. Therefore, in this example a very bad gauge with a rank $\sim$100 could take a significantly large computational time. } }
\label{fig:PiElite_vs_Performance}
\end{figure*}

Benchmark studies assessing the presence or absence of speed-up of quantum annealers compared to classical processors resort to gauge selection as a way of obtaining reliable averaged results of the performance of the device~\cite{boixo_evidence_2014,Ronnow25072014,King_arXiv2014,PerdomoOrtiz_EPJST2015}. Although it is known that gauge specification can significantly enhance the performance of the device, previous studies are limited to the scaling of the typical gauge since there is no \textit{a priori} way to determine the optimal gauge. Gauge specification is a particular example of Hamiltonian specification discussed above. We present here how our performance estimator $\Pi^{\epsilon\%}_{\rm{elite}}$ can be used to select the optimal gauges.
To illustrate the procedure, we used a hard-instance out of a pool of random-spin glass instances similar to the ones reported elsewhere~\cite{Ronnow25072014}. This instance was provided by Sergio Boixo, who assessed that this particular instance had a simulated-annealing (SA) runtime of the order of hundred times longer than the median instance, from a pool of 	hundreds of thousands of instances within the same family (instances with random couplings $J_{ij} \in \{+1,-1\}$, with 509 qubits as shown in Fig.~\ref{fig:chimera}). As an abbreviation, we refer hereafter to this specific random-spin glass example as the RS instance. For QA this instance was shown to also be particularly hard after not obtaining any ground states after trying 16 gauges, with 10,000 readouts each, at 20 $\mu$s. Fig.~\ref{fig:PiElite_vs_Performance}(b) corroborates this assessment; the median gauge over a set of 100 random gauges has a $p_s  =1/(2\times10^6) = 5\times 10^{-7}$, resulting in a expected number of repetitions to solution of $\bar{R}_{.99} \sim 9.2 \times 10^{6}$ annealing cycles. Since $\bar{R}_{.99}$ is about two hundred times greater than $N_{\rm{reads}} = 50,000$, applying our performance estimator to select the optimal gauge before engaging in lengthier runs is expected to significantly reduce the computational time. 

Fig.~\ref{fig:PiElite_vs_Performance}(a) shows there is a strong correlation between the rank obtained with $\Pi^{2\%}_{\rm{elite}}$ with $N_{\rm{reads}} = 50,000$ and $n_{\rm{reps}} = 1$, compared to the greedy performance rank described above. The number of total readouts used to estimate the performance rank is $2$ million per gauge. The error bars correspond to the rank provided by the first and third quartile out of 40 different experiment each with $N_{\rm{reads}} = 50,000$ for each of the 100 random gauges. The middle point corresponds to the median of the set of experiments. Fig.~\ref{fig:PiElite_vs_Performance}(b) shows the same data set but with the raw values for $\Pi^{2\%}_{\rm{elite}}$ and also serves the purpose of showing the count in the number of ground states, illustrating that gauge selection can have a significant impact in the device performance, an increase as much as one to two orders of magnitude (see also Fig.~\ref{fig:PiElite_vs_Performance}(d) and Fig.~\ref{fig:consist_6F_20us}(b) reflecting how the gauge selection can influence $n_{\rm{gs}}$.)

As expected, any performance estimator would be a noisy metric and not expected to have a 100\% correlation with the real performance from an extensive number of readouts $N_{\rm{total}} \sim R_{.99}$. The RS section of Table~\ref{table:fractions_greedyrank} (upper half) addresses this issue. 
Suppose one utilizes the following strategy. One decides to run 100 gauges with a fixed $N_{\rm{reads}}$ for each gauge. From this starting data set, and while processing the readouts in the search of an optimal solution, one can easily calculate $\Pi^{\epsilon \%}_{\rm{elite}}$. Since $N_{\rm{reads}} \ll R_{.99}$, it is unlikely that this initial batch of calculations would contain the optimal solution. The refinement we propose here consists of using the information of the $\Pi^{\epsilon \%}_{\rm{elite}}$ calculated \textit{on-the-fly} to select, for example, the top 5 gauges (gauges with the highest $\Pi^{\epsilon \%}_{\rm{elite}}$ score) out of the 100 random gauges. Since the selected gauges are expected to have a better performance than the typical or average gauge, only the selected ones  are used to continue with the remaining runs $N_{\rm{total}}\sim R_{.99}$ until the desired solution is found.  Given this strategy, Table~\ref{table:fractions_greedyrank} answer the question: what is the probability that the absolute top gauge (that is, ranked number 1 according to the performance rank in Sec.~\ref{subsec:ranking}) is contained in this set of predicted top 5 gauges? What is the probability of one finding any of the top 2 gauges in the set of predicted top 5 gauges? etc, etc. Notice that at the level of $N_{\rm{reads}} =  50,000$ which is $\sim$200 times less than $\bar{R}_{.99}$, one obtains a reasonably high $\sim$75\% probability of obtaining the top 1 gauge in the set top 5 gauges predicted by $\Pi^{2 \%}_{\rm{elite}}$. Table~\ref{table:fractions_greedyrank} also addresses the question of the existence of an optimal value of $\epsilon$ for the performance estimator. In all the examples considered here it seems to be the case that a value of $\epsilon\% = 1$\% or $2$\% is optimal, a non-trivial result, since one might think incorrectly that the greater the number of low energy states included in the calculation of $\Pi^{2 \%}_{\rm{elite}}$ the better. This table also shows the expected increase in the probability of choosing the top gauges as $N_{\rm{reads}}$ becomes larger.
Note also the inclusion in the table of a ``Greedy" column for each case. This new metric was included because the use of the greedy method for the \textit{performance rank} described in Section B begs the question of why one could not use an even simpler performance estimator consisting of the same \textit{greedy approach} applied here to the case of a small number $N_{\rm{reads}}$ instead of $N_{\rm{total}}$. However, the table clearly shows that $\Pi^{2 \%}_{\rm{elite}}$ is consistently always as good as, if not much better than, the greedy performance estimator. This is not surprising since as expected the $\Pi^{2 \%}_{\rm{elite}}$ should be a more robust metric than the greedy approach and to be less sensitive to rare-event occurrences. A clear advantage of the $\Pi^{2 \%}_{\rm{elite}}$ is that it provides a score function that can be used for other purposes as will be shown elsewhere~\cite{PerdomoOrtiz_arXiv2015b}. The greedy approach as a score function is expected to be much flatter, due to this ranking relying heavily on the breaking of ties whenever there are only a few low energy states that many gauges reach).

%\begingroup{
\begin{table*}[!t]
\tiny
%\renewcommand{\arraystretch}{1.3}
\caption{\small{Fraction of experiments where the performance estimator $\Pi^{\epsilon\%}_{\rm{elite}}(N_{\rm{reads}})$ predicted correctly the top 1, either the top 1 or 2 (labeled Any Top 2), and any among the top 1, 2, or 3 (labeled Any Top 3), etc, within the set of 5 top-ranked gauges out of 100 random gauges, for the cases of the random-spin (RS) and the hardest fault diagnosis  (50M-DMF) problem instances described in the main text.  The number of experiments is 400, 200, 80, and 40 for $N_{\rm{reads}} = 5,000, 10,000, 25,000$ and $50,000$, respectively. Note the non-trivial dependence on the fraction included in the elite mean, considering values of $\epsilon  = 1, 2, 5$, and $10$ \%, and showing an intermediate optimal value around 1\% or 2\%. As explained in the text, the \textit{greedy} approach corresponds to the (1/$N_{\rm{reads}}$)*100-percentile, i.e., rank based on lowest energy obtained and breaking ties with the frequency of the lowest energy states). Annealing time per readout for all experiments was 20 $\mu$s.}}
\label{table:fractions_greedyrank}
\centering
% Some packages, such as MDW tools, offer better commands for making tables
% than the plain LaTeX2e tabular which is used here.
\begin{ruledtabular}
\begin{tabular}{c|ccccc|ccccc|ccccc|ccccc|ccccc}
\textbf{RS} & \multicolumn{5}{c|}{Top 1} & \multicolumn{5}{c|}{Any Top 2} & \multicolumn{5}{c|}{Any Top 3} & \multicolumn{5}{c|}{Any Top 4} & \multicolumn{5}{c}{Any Top 5} \\
\hline
$N_{\rm{reads}}$ & Greedy & 1\% &  2\%  & 5\%  & 10\% & Greedy & 1\% &  2\%  & 5\%  & 10\% & Greedy & 1\% &  2\%  & 5\%  & 10\% & Greedy & 1\% &  2\%  & 5\%  & 10\%  & Greedy & 1\% &  2\%  & 5\%  & 10\% \\
\hline
5k & 0.50 & 0.52 &  0.53  & 0.53  & 0.41 & 0.64 & 0.73 &  0.73 & 0.66  & 0.52 & 0.88 & 0.94 &  0.95  & 0.93  & 0.81 & 0.89	&0.91	&0.91	&0.9	0 & 0.86 & 0.97	& 0.97	& 0.97	& 0.97	& 0.94\\
10k & 0.53 & 0.57 &  0.57  & 0.57  & 0.43 & 0.67 & 0.70 &  0.73 & 0.69  & 0.59 & 0.90 & 0.92 &  0.92  & 0.92 & 0.84 & 0.93	& 0.94	&  0.94	& 0.92 &	0.84 & 0.99	& 0.99	& 0.99	& 0.99	& 0.96 \\
25k & 0.58 & 0.64 &  0.64  & 0.66  & 0.45 & 0.70 & 0.74 &  0.75 & 0.75  & 0.60 & 0.83 & 0.88 &  0.88  & 0.95  & 0.85 & 0.9	& 0.91	&  0.9	& 0.95 &	 0.85 & 0.96	& 0.98	& 0.98	& 1.00	& 0.98\\
50k & 0.73 & 0.75 &  0.75  & 0.68  & 0.50 & 0.88 & 0.88 &  0.88 & 0.75  & 0.60 & 0.95 & 0.95 &  0.95  & 0.95  & 0.93 & 1.00 & 	0.95 & 	0.95 & 	0.95&	0.93 & 1.00 & 	1.00	& 1.00	& 1.00 & 	1.00 \\
\hline\hline 
\textbf{50M-DMF} & \multicolumn{5}{c|}{Top 1} & \multicolumn{5}{c|}{Any Top 2} & \multicolumn{5}{c|}{Any Top 3} & \multicolumn{5}{c|}{Any Top 4} & \multicolumn{5}{c}{Any Top 5} \\
\hline
$N_{\rm{reads}}$ & Greedy & 1\% &  2\%  & 5\%  & 10\% & Greedy & 1\% &  2\%  & 5\%  & 10\% & Greedy & 1\% &  2\%  & 5\%  & 10\% & Greedy & 1\% &  2\%  & 5\%  & 10\%  & Greedy & 1\% &  2\%  & 5\%  & 10\% \\
\hline
5k & 0.49	& 0.58 & 	0.59&	 0.57& 	0.54 & 0.68&	 0.78& 	0.78	 & 0.73	& 0.68 & 0.73 &	0.83	 &  0.83 & 	0.78	 & 0.74 & 0.73 	& 0.84	 & 0.83	& 0.79 &	0.75 & 0.86	& 0.93 	& 0.93 &	0.87	 & 0.82\\
10k & 0.54	& 0.64 & 	0.64	 & 0.62	& 0.6 & 0.72 & 	0.8	& 0.81	& 0.78 	& 0.74 & 0.77	& 0.86 & 	0.87	 & 0.83	& 0.79 & 0.77	& 0.86	&  0.87	& 0.83 	& 0.79 & 0.89	& 0.95 &	 0.95 &	0.9	& 0.85 \\
25k & 0.62	& 0.72	& 0.74	& 0.72 & 	0.72 & 0.85	& 0.85 	& 0.85 & 	0.82 &	0.8 & 0.90	& 0.91 &	 0.9	 & 0.86	& 0.85 & 0.90	&  0.91	& 0.9	& 0.86 	& 0.85 & 0.98	& 0.97	& 0.98 &	0.91	 & 0.91\\
50k & 0.71	& 0.84	& 0.82	& 0.75 & 	0.73 & 0.92	& 0.92 & 	0.88 	& 0.85 & 	0.84 & 0.93	& 0.95	&  0.91	& 0.9	& 0.87 & 0.93	& 0.95	 & 0.91 & 	0.9	& 0.87 & 0.99	& 0.99	 & 0.97	& 0.95	& 0.91 \\
\end{tabular}
\end{ruledtabular}
\end{table*}
%}

\subsection{Iterative strategy for embedding parameter setting and gauge selection}\label{subsec:tuneJE}

When solving real-world applications there are several additional subtleties to take into consideration. While in the case of the random spin-glass instances there is only one objective function appearing in the quantum annealing implementation, $E_{\rm{ising}}$ in Eq.~\ref{eq:Eising}, in instances derived from real-world applications $E_{\rm{ising}}$ is obtained from another cost energy function $E_{\rm{QUBO}}$, a quadratic binary expression containing the logical qubits, $\{ \boldsymbol{\vec{s}}_{\rm{logical}} \}$ before these are embedded into the hardware qubits, $\{ \boldsymbol{\vec{s}}_{\rm{hardware}} \}$ appearing in $E_{\rm{ising}}$~\cite{PerdomoOrtiz_EPJST2015}. From a practical application perspective, we are interested in the possibility of using our performance estimator to select the best Hamiltonian specification with the smallest $R_{.99}$, therefore reducing the computational time.

\begin{figure}
\centering
\includegraphics[width=0.48\textwidth]{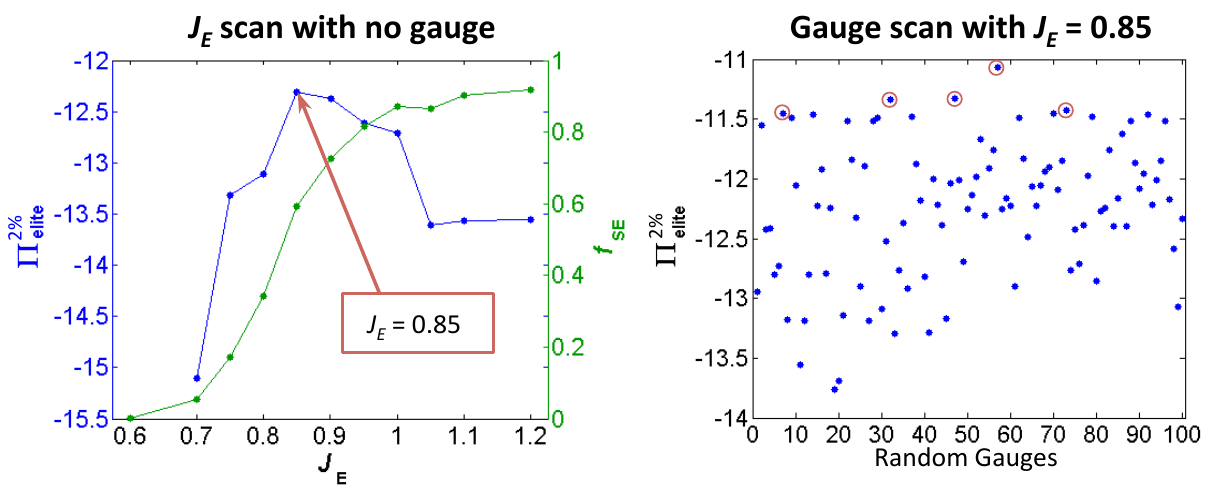}
\caption{\scriptsize{ \textbf{Iterative strategy for parameter setting and gauge selection for the 50M-DMF instance:} The corresponding data illustrates only one realization for the selection of $J_E$ (left) and another for the selection of the top 5 gauges (circled) with the highest score (right). For a statistical analysis of the robustness of the method, see the second part of Table~\ref{table:fractions_greedyrank}. The left y-axis for the $J_E$ plots the value of the $\Pi^{2\%}_{\rm{elite}}$ when calculated using $N_{\rm{reads}} = 50,000$ which is more than three-orders of magnitude less than the number of readouts to solution, $R_{.99}$, for this problem. The right y-axis for this $J_E$ plot shows the percentage of solutions that passed strict embedding (solutions with no violations of the constraints imposed by $J_E$). $f_{\rm{SE}}$ serves as a guide for selecting the region with the optimal $J_E$, as explained in the main text.}}
\label{fig:cycle_Needle_20us}
\end{figure}

Once the embedding problem is solved~\cite{Choi2008,Choi2011,Cai-14} and one has a mapping of each logical qubit $s_{i,l} \in \{ \boldsymbol{\vec{s}}_{\rm{logical}} \}$ into a subset of $\{ \boldsymbol{\vec{s}}_{\rm{hardware}} \}$, the first decision to be made is to select the strength of the coupling $J_E$ needed to keep the hardware spins representing a logical spin in alignment with each other.~\footnote{For simplicity we assume that all these penalties $J_E$ enforcing the embedding are equal. Further fine tuning can be done by optimizing each parameter but this is beyond the scope of this work.} For each value of $J_E$, and after requesting $N_{\rm{reads}}$, we can calculate the fractions of the $N_{\rm{reads}}$ that do not violate any of these embedding constraints, i.e., with all the physical spins representing logical spins being properly aligned. These solutions are said to satisfy the \textit{strict embedding} (SE) requirement. The right $y$-axis in Fig.~\ref{fig:cycle_Needle_20us}(a) shows the fraction of solutions passing this requirement out of the total number of readouts $N_{\rm{reads}}$, denoted as $f_{\rm{SE}}(J_E)$. 

Intuitively the magnitude of $J_E$ cannot be too weak since it will not achieve the goal of keeping the spins properly aligned (same readout value for each variable representing the same logical qubit). Having a large value $J_E$ will certainly help in increasing the probability of not having any misaligned spins but it cannot be too strong either, $J_E \gg \max(J_{ij})$, since after dividing everything by $J_E$ to make all $-2 \le h_i \le 2$ and $-1 \le J_E, J_{ij} \le 1$, the original values $h_i$ and $J_{ij}$ will be well below the precision level and the performance will be significantly affected by noise. Therefore, a sweet spot with an optimal value of $J_E$ is expected. From our experience, $f_{\rm{SE}}$ serves as a guide for selecting the region of interest, denoted here as $J^{*}_E < J_E < J^{**}_E$, with $f_{\rm{SE}}(J^{*}_E) \approx 0.05$ and $J^{**}_E$  corresponding to the onset of the plateou region with $f_{\rm{SE}} \approx f^{\rm{max}}_{\rm{SE}}$ in the plot $f_{\rm{SE}}$ vs. $J_E$. The value $f^{\rm{max}}_{\rm{SE}}$ can be easily obtained experimentally in one-shot by setting $J_E \gg 1$, and one can use that value to search for $J^{**}_E$.

%It is a big challenge to obtain a sizable number of ground states. If possible, one could calculate $p_s$ for every single gauge and every value of $J_E$ (Hamiltonian specification) and therefore be able to rank them by this gold-standard performance rank. As shown in Fig.~\ref{fig:PiElite_vs_Performance}(b,d), even after $N_{total}=2$ millions and 5 millions for the RS and 50M-DMF, respectively, there were only a handful of gauges with $n_{\rm{gs}} \ge 1$ (this difficulty being the main motivation for defining the \textit{greedy performance rank} in Sec.~\ref{subsec:ranking}). 

\begin{figure*}
\centering
\includegraphics[width=0.95\textwidth]{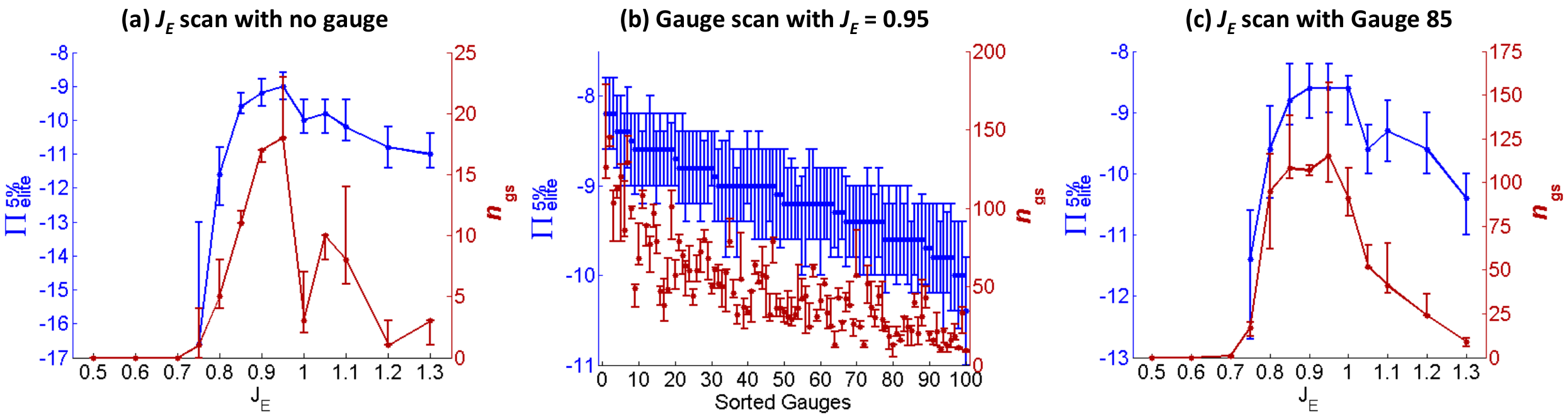}
\caption{\scriptsize{The plots showing the strong correlation between our score function, $\Pi^{5\%}_{\rm{elite}}$, and number of ground states, $n_{\rm{gs}}$, for the 300K-DMF problem instance. The number of readouts, $N_{\rm{reads}}=100$, used to compute the score function is at least two orders of magnitudes less than the estimation of the typical number of readouts needed to observe the optimal solution at least once with a probability of 99\% ($R_{.99} = 10,707$ for the median performance over 100 random gauges). The error bars in $\Pi^{5\%}_{\rm{elite}}$ (blue) corresponds to the 25-percentile and the 75-percentile from 500 experimental realizations. The error bars in $n_{\rm{gs}}$ (red) corresponds to the lowest and highest of three realizations, each with $N_{\rm{total}}= 100,000$, with the middle value as the median of the three experimental realizations. This shows that in these hard-instances, $n_{\rm{gs}}$ is still a noisy value, but still our performance estimator places us in the range of the top gauges with the largest $n_{\rm{gs}}$. Each experimental realization consists of the estimation of $\Pi^{5\%}_{\rm{elite}}$ ($n_{\rm{gs}}$) by using $N_{\rm{reads}}$ ($N_{\rm{total}}$) per gauge. The iterative approach described here is the same one described in the text and in Fig.~\ref{fig:cycle_Needle_20us} to set the embedding parameter $J_E$ and to select the optimal gauges.}}
\label{fig:consist_6F_20us}
\end{figure*}

For the purpose of our discussion we selected two instances from the fault diagnosis application published elsewhere~\cite{PerdomoOrtiz_EPJST2015}. The first instance, referred hereafter as 300K-DMF, was selected because despite its implementation with only 81 hardware qubits, it is unusually hard\footnote{For the 300K-DMF, median success probability of $\bar{p}_s = 129/300,000 = 4.3/10,000$ out of set of 100 random gauges} when compared with other benchmark studies~\cite{Ronnow25072014}, yet has a success probability just high enough to  allow for a sizable number of ground states even in the worst Hamiltonian specification,\footnote{For the 300K-DMF, the smallest $p_s =25/300,000 = 8.3/100,000$} within a reasonable number of readouts set to $N_{\rm{reads}} = 300,000$ per gauge or $J_E$ considered. As shown in Fig.~\ref{fig:consist_6F_20us}, a finite number of ground states for every single gauge and every value of $J_E$ allows to rank each of these Hamiltonian specifications by their gold-standard performance rank and to compare with the rank predicted from our performance estimator. To satisfy the condition $\bar{R}_{.99} \gg N_{\rm{reads}}$, the $\Pi^{5\%}_{\rm{elite}}$ per gauges was calculated by using only $N_{\rm{reads}} =  100$. Fig.~\ref{fig:consist_6F_20us}, shows that even with so few readouts, there is a strong correlation between our $\Pi_{\rm{elite}}$ score function and the number of ground states observed after $N_{\rm{total}} = 300,000$. Compared to the harder instances where $N_{\rm{reads}} = 50,000$, here we used an $\epsilon \%$ of 5\% instead of 2\%, since the latter would amount to computing the elite mean with only the two lowest values out of $N_{\rm{reads}}=100$. This makes the estimator too noisy and flat, analogous to the ``Greedy" performance estimator that only uses the lowest value to rank gauges. As shown in Fig.~\ref{fig:consist_6F_20us}, calculating the elite mean over the five lowest energies already gives a good correlation with $n_{\rm{gs}}$.

The second instance, referred hereafter as 50M-DMF instance, serves the purpose of showing how our performance estimator can be used in a practical situation, for instances with probabilities much smaller. These are the instances we expect to surface in the next generation of quantum annealers. The instance 50M-DMF has the property of having a unique optimal solution, making it the most difficult to solve among the family of problem instances with six-faults to be diagnosed. More specifically, although the number of hardware qubits (96 qubits) required to implement this instance is not unusually large, this instance turned out to be extremely difficult for QA; not even a single-ground state was measured after $N_{\rm{total}} = 50 \times 10^6$ annealing cycles, even after optimizing for the optimal $J_E$ but under the default no-gauge! This instance was in large the motivation for defining a quick strategy to find the optimal Hamiltonian specifications (best $J_E$ and best gauge) capable of finding the ground state).

Fig.~\ref{fig:cycle_Needle_20us} describes the suggested iterative approach used to optimize both the value of $J_E$ and to select the optimal gauges in instances requiring a direct embedding approach. Starting with the no-gauge one can scan for the candidate values of $J_E$ and select the value of $J_E$ with the highest score $\Pi^{2\%}_{\rm{elite}}$. In contrast with the case of the RS instance above, here we cannot use $E_{\rm{ising}}$ to calculate the score function since, for example, the lowest energies of $E_{\rm{ising}}$ will be different for every value of $J_{E}$ (because of the energy renormalization to fit all programmable values $h_i, J_{ij},$ and $J_E$ within the dynamical range $\abs{h_{i}} \le 2$ and $\abs{J_{ij}} \le 1$). To circumvent this issue we compute $\Pi^{2\%}_{\rm{elite}}$ after error-correcting the $N_{\rm{reads}}$ solutions with majority voting~\cite{King_arXiv2014} when going from  $\{ \boldsymbol{\vec{s}}_{\rm{hardware}} \} \rightarrow \{ \boldsymbol{\vec{s}}_{\rm{logical}} \}$, and then sorting the states according to $E_{\rm{QUBO}}(\{ \boldsymbol{\vec{s}}_{\rm{logical}} \})$ before selecting the 2\% of the lowest energies used in the computation of $\Pi^{2\%}_{\rm{elite}}$.

The next step in the procedure is to perform a \textit{gauge scan} at the selected value of $J_E$ from the $J_E$ \textit{scans}. Considering that we are dealing with instances with $R_{.99} \gg N_{\rm{reads}}$, it is not a significant usage of computational resources of perform calculations with a number of gauges on the order of about 100. Notice that there is really no overhead while doing the gauge scans, since for every gauge considered, one needs to post-process all the solution readouts (e.g., with majority voting) while searching for the states with the optimal solutions anyways. Since the energies of every single solution needs to be calculated, the only overhead in calculating $\Pi^{2\%}_{\rm{elite}}$ comes from a cheap sorting of these energies before calculating the elite mean. For NP-complete problems, we can always tell if we have found the desired answer. Also, for a large family of problems such as those NP-hard problems where the NP-complete version is still interesting, one can still stop the search if the desired solution is obtained (e.g., we can ask whether there exists a solution with an energy lower than a reference energy, with the latter being for example the best solution attainable with a state of the art classical solver). Therefore, trying $\sim$100 gauges in the search of an optimal gauge is not an unfeasible idea. 
After calculating $\Pi^{2\%}_{\rm{elite}}$ for the complete pool of gauge candidates, one can proceed to another set of $J_E$-scans by using the best gauge with the highest score. As shown in Fig.~\ref{fig:consist_6F_20us}(c) and from our experience with other problem instances where the procedure was even applied at different annealing times, in most of the cases the second optimal $J_E$ matched the same $J_E$ from the first scan under the no-gauge setting. Even in the cases where $J_E$ moved to a new value, the change was in the neighborhood of the first optimal value. As shown in Fig.~\ref{fig:consist_6F_20us}(a) and (c), as long as one is near the optimal value of $J_E$ the performance is not significantly affected. The gauge selection, Fig.~\ref{fig:consist_6F_20us}(b), seems to have a much larger impact in the performance. Since the first $J_E$-scan does the job of taking us to the neighborhood of $J_E$ optimality (out of the set of candidates considered), it is reasonable to conclude that a second $J_E$ is not necessary and it is better to focus on the top gauges obtained from a gauge scan of 50-100 gauges. For easy instances a large gauge set would be unnecessary since the optimal solution will likely appear before one finishes going through the target number of 100 or so gauges.

As shown in Fig.~\ref{fig:PiElite_vs_Performance}, the performance estimator proposed here is a noisy metric. For example, there is no guarantee that the top gauge is the same one as the one predicted by $\Pi^{2\%}_{\rm{elite}}$. Therefore, instead of selecting only the gauge predicted as top 1, it is advisable to select a handful of the predicted as top gauges as indicated in Fig~\ref{fig:cycle_Needle_20us}(b). It is with this selected set that one performs the extensive runs $N_{\rm{total}}\sim R_{.99}$, but where now $R_{.99}$ has been significantly reduced given that we are running with a set that includes the optimal gauge from the random set. Table~\ref{table:fractions_greedyrank} shows that selecting the predicted top five gauges has a high probability $(>80\%)$ of containing the top 1 gauge yielding the largest number of ground states. Predicting any of the top 2 gauges had a probability $\sim 90\%$. This is very remarkable considering that in this particular 50M-DMF problem running a low-performing gauge would lead to a significantly large time to solution. As mentioned above, the default no-gauge did not find the solution after 50 millions reads, therefore yielding a $R_{.99} > 230$ millions, while any of the top 3 gauges require $R_{.99} < 23$
millions, providing at least an order of magnitude improvement in this hard-to-solve instance for the QA processor.

In the case of real-world applications that use ancilla variables~\cite{Babbush2014,PerdomoOrtiz_EPJST2015} in the construction of their $E_{\rm{QUBO}}$ post-processing strategies are also possible. In these cases it is more efficient to process the solution, for example, evaluating the problem energy, $E_{\rm{problem}}$ with only the relevant variables defining the problem. More specifically, and without loss of generality, we can express the set of resulting logical qubits in the $E_{\rm{QUBO}}$ expression as $\{ \boldsymbol{\vec{s}}_{\rm{logical}} \}$  = $\{\boldsymbol{\vec{s}}_{\rm{problem}} \} \cup \{ \boldsymbol{\vec{s}}_{\rm{ancilla}} \}$, where $\{\boldsymbol{\vec{s}}_{\rm{problem}} \}$ corresponds to the set of qubits or binary variables that define completely the problem description and that can be extracted to evaluate the energy of the problem, $E_{\rm{problem}}$. In these cases, and with the intention of increasing the chances of finding the optimal solution, it is more efficient to process the solution readouts with $E_{\rm{problem}}(\{\boldsymbol{\vec{s}}_{\rm{problem}} \})$ and not with $E_{\rm{QUBO}}(\{\boldsymbol{\vec{s}}_{\rm{logical}} \})$. This postprocessing strategy can only help in finding the optimal solution, since for every readout $E_{\rm{problem}}(\{\boldsymbol{\vec{s}}_{\rm{problem}} \}) \le E_{\rm{QUBO}}(\{\boldsymbol{\vec{s}}_{\rm{logical}} \})$, therefore allowing for the possibiliy of finding optimal solutions in solutions that had been penalized by the ancilla constrains. Our preliminary results indicate that for these problem instances, it is advisable to look also at the top 5 gauges obtained from the greedy approach (same approach described in Table~\ref{table:fractions_greedyrank} but now using $E_{\rm{problem}}$ instead of $E_{\rm{QUBO}}$) along with the top 5 from the $\Pi_{\rm{elite}}$ score function, also calculated with $E_{\rm{problem}}$ instead of $E_{\rm{QUBO}}$. The strategy proposed here consists of taking as the ``selected top gauges" the union set of these two sets of top 5 gauges, and perform with these gauges the extensive runs with $N_{\rm{total}} \sim R_{.99}$. 

\section{Conclusions}\label{sec:conclusions}

We defined a score function intended to estimate the performance of quantum annealers whose applicability does not rely on obtaining ground states corresponding to the desired solution. We observed a strong correlation of our performance estimator with the performance of the device even in the case where the number of readouts used to calculate it was several orders of magnitude less than the number of readouts required to find the desired solutions. The score function is based on a tail conditional expectation value, corresponding to the \textit{elite mean} over a small percent representing the readouts with the lowest energies. We showed it can be used to efficiently select the optimal gauges from a large pool of random gauges and in setting Hamiltonian parameters appearing in the implementation of real-world applications.

Although it has been previously shown that the decisions in programming quantum annealing devices can significantly impact the performance of the device~\cite{boixo_evidence_2014,PerdomoOrtiz_EPJST2015,King_arXiv2015}, thus far comparison of performance of quantum annealers to algorithms on conventional classical processors was limited to average performance over the selection of parameters explored. This study opens the possibility of revisiting such scaling studies, now with the opportunity to select in advance the best configuration of the device. Having the possibility of selecting the specifications (best gauges or other optimal parameters) will be indispensable once we start solving instances intrinsically harder with the new generation of quantum annealers. The overhead incurred to apply the selection procedure presented here is constant and it does not scale with the size of the system. We showed that even in cases where $N_{\rm{reads}} < R_{.99}/1000$, the method still works with a large probability of selecting the top gauges. 

In the case of real-world applications, the iterative strategy proposed in Sec.~\ref{subsec:tuneJE} requires essentially no overhead in calculating the performance estimator and used to rank the random gauges. Since the data needs to be processed anyway (e.g., calculation of $E_{\rm{QUBO}}$ and majority voting while testing whether or not the desired solution has been found), the only overhead incurred is the time needed to sort the solution before calculating the elite mean. In the case of other parameter settings such as the one used in the embedding problem, our performance estimator provides a very efficient approach by pinning down the region where the device has its best performance. 

Although our strategy allows us to select the best Hamiltonian specification in quantum annealers, we do not expect that it will be enough to change the complexity class seen in scaling studies~\cite{boixo_evidence_2014,Ronnow25072014}. Certainly, it could easily provide a speed-up of an order of magnitude from the default methods, as seen in some of the examples presented here, and it might be to-date the only feasible way to obtain solutions to hard-to-solve computational problems (either random spin-glass benchmarks or real-world applications) in the next generations of quantum annealers.

\section*{Acknowledgements}

This work was supported in part by the Office of the Director of National Intelligence (ODNI), Intelligence Advanced Research Projects Activity (IARPA), via IAA 145483. We want to acknowledge the support of NASA Advanced Exploration Systems program and NASA Ames Research Center. The authors thank Sergio Boixo for providing the 509-qubit RS problem instance, and Bryan O'Gorman, Eleanor Rieffel, and Davide Venturelli for helpful discussions. 

\section*{Author contributions}
A.P-O, J.F, R.B and V.N.S contributed to the ideas presented in the paper. A.P-O and J.F designed and ran the experiments and wrote the manuscript. All the authors revised the manuscript.

%\subsection*{Competing financial interests}
%
%The authors declare no competing financial interests.

\appendix 

\clearpage
\section{Chimera architecture}\label{sec:progQA}
\begin{figure}[!h]
\centering
\includegraphics[width=0.45\textwidth]{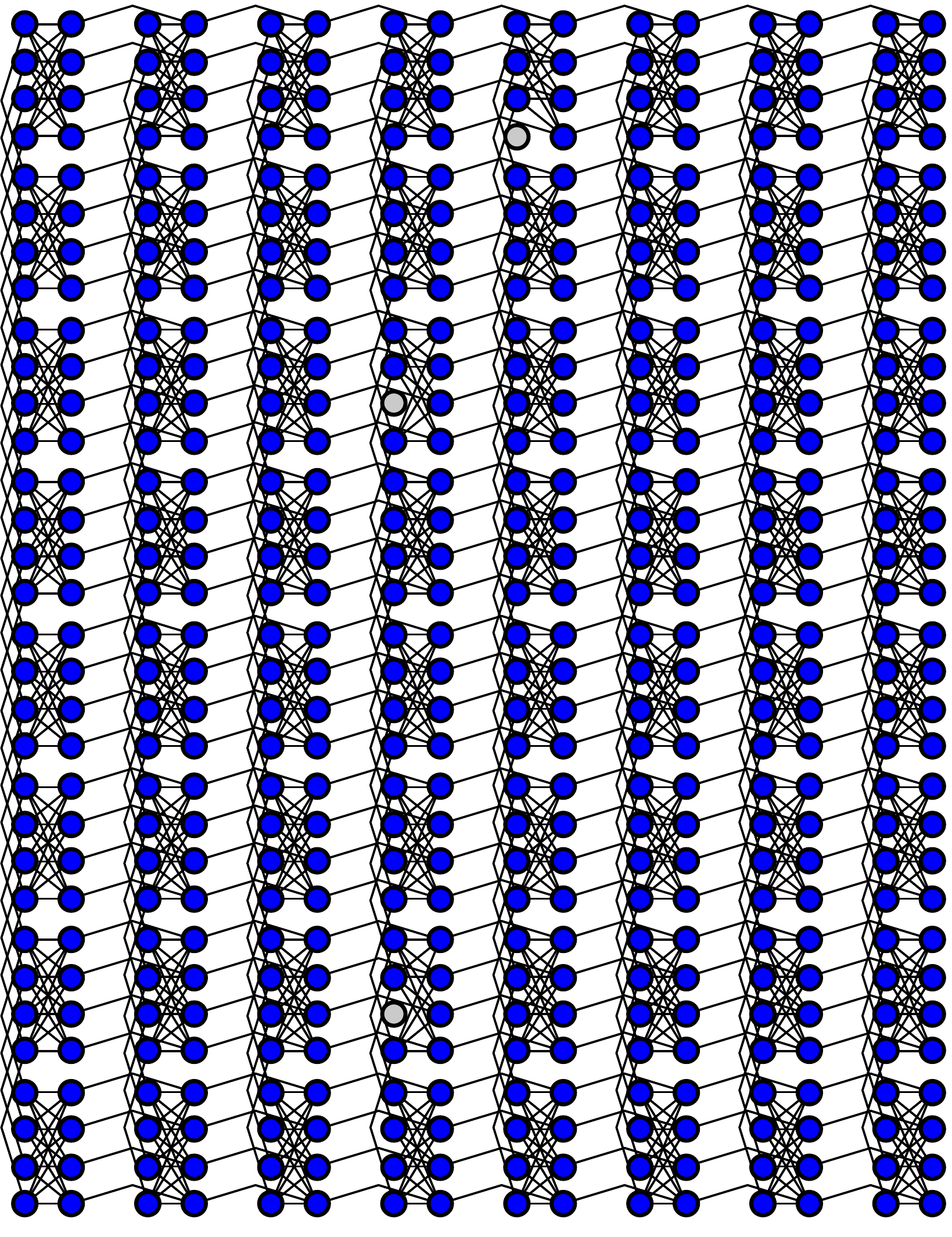}
\caption{\textbf{Device architecture and qubit connectivity D-wave Two at NASA Ames:}. The array of superconducting quantum bits is arranged in $8\times 8$ unit cells that consist of 8 quantum bits each. Within a unit cell, each of the 4 qubits in the left-hand partition (LHP) connects to all 4 qubits in the right-hand partition (RHP), and vice versa. A qubit in the LHP (RHP) also connects to the corresponding qubit in the LHP (RHP) of the units cells above and below (to the left and right of) it. Edges between qubits represent couplers with programmable coupling strengths. Blue qubits indicate the 509 usable qubits, while grey qubits indicate the three unavailable ones out of the 512 qubit array.}
\label{fig:chimera}
\end{figure}

\section{Evidence against a commonly used rule-of-thumb for gauge selection}\label{app:thumbrule}

\begin{figure}[!h]
\centering
\includegraphics[width=0.48\textwidth]{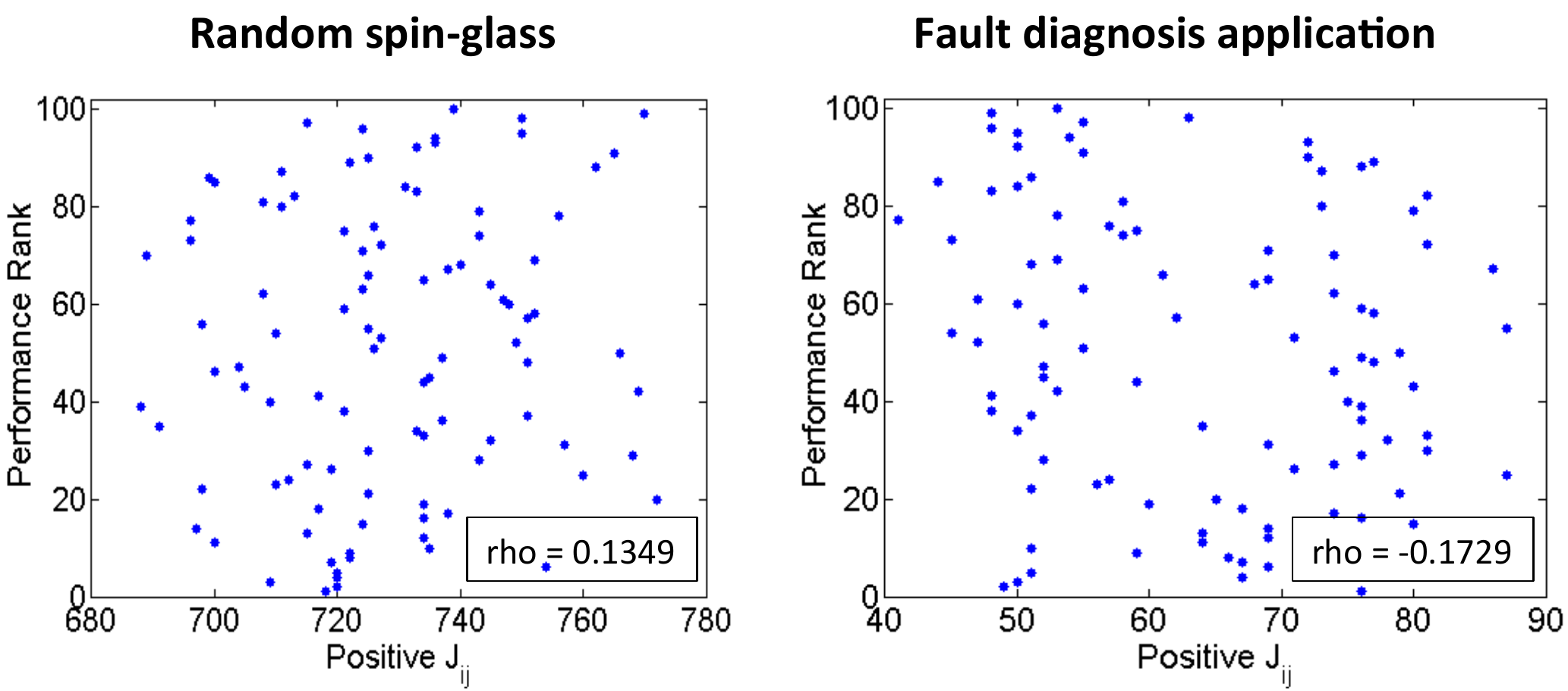}
\caption{The low value of spearman correlation coefficient shows that there is basically no correlation between the number of $J_{ij}>0$ resulting from a specified gauge and the performance in the device, ruling out the common belief that the larger the number of $J_{ij} > 0$, the better. Shown here are examples from three different application domains. }
\label{fig:Jpositives}
\end{figure}

In the case of gauge selection, a commonly used "rule-of-thumb" that had persisted in the community is that the gauge maximizing the number of antiferromagnetic couplings, $J_{ij}>0$ is preferred. The physical motivation behind this ``rule" is that the precision in the specification of a $J_{ij}>0$ (antiferromagnetic coupling) is more robust than its negative (ferromagnetic) counterpart~\cite{TrevorJposComment}. A more detailed analysis including 100 gauges for several problem applications considered (see Fig.~\ref{fig:Jpositives}) shows that such rule-of-thumb does not hold in any of the hard instances considered here. Notice there is no correlation between the number of positive couplers  and the performance of the specified gauge. We did no see any correlation either in any other quantity similar to $J_{ij}$: parameters studied include the number of $J_E>0$ (for the case of real-world applications with direct embedding), the number of $h_i >0$ and the number of $J_{ij}$ that are non-$J_E$. For all those cases, still no correlation was found.

\begin{widetext}

\end{widetext}

%% Create the reference section using BibTeX:
%\bibliographystyle{naturemag}
%%\bibliography{refsAQCall,space}
%\bibliographystyle{naturemag}
%\bibliography{refsAQCall_03032015}

\begin{thebibliography}{10}
\expandafter\ifx\csname url\endcsname\relax
  \def\url#1{\texttt{#1}}\fi
\expandafter\ifx\csname urlprefix\endcsname\relax\def\urlprefix{URL }\fi
\providecommand{\bibinfo}[2]{#2}
\providecommand{\eprint}[2][]{\url{#2}}

\bibitem{Bunyk_IEEE2014}
\bibinfo{author}{Bunyk, P.} \emph{et~al.}
\newblock \bibinfo{title}{Architectural considerations in the design of a
  superconducting quantum annealing processor}.
\newblock \emph{\bibinfo{journal}{Applied Superconductivity, IEEE Transactions
  on}} \textbf{\bibinfo{volume}{24}}, \bibinfo{pages}{1--10}
  (\bibinfo{year}{2014}).

\bibitem{johnson_quantum_2011}
\bibinfo{author}{Johnson, M.~W.} \emph{et~al.}
\newblock \bibinfo{title}{Quantum annealing with manufactured spins}.
\newblock \emph{\bibinfo{journal}{Nature}} \textbf{\bibinfo{volume}{473}},
  \bibinfo{pages}{194--198} (\bibinfo{year}{2011}).

\bibitem{kadowaki_quantum_1998}
\bibinfo{author}{Kadowaki, T.} \& \bibinfo{author}{Nishimori, H.}
\newblock \bibinfo{title}{Quantum annealing in the transverse ising model}.
\newblock \emph{\bibinfo{journal}{Phys. Rev. E.}}
  \textbf{\bibinfo{volume}{58}}, \bibinfo{pages}{5355} (\bibinfo{year}{1998}).

\bibitem{Farhi2001}
\bibinfo{author}{Farhi, E.} \emph{et~al.}
\newblock \bibinfo{title}{A quantum adiabatic evolution algorithm applied to
  random instances of an {NP-Complete} problem}.
\newblock \emph{\bibinfo{journal}{Science}} \textbf{\bibinfo{volume}{292}},
  \bibinfo{pages}{472--475} (\bibinfo{year}{2001}).

\bibitem{PerdomoOrtiz2012_LPF}
\bibinfo{author}{Perdomo-Ortiz, A.}, \bibinfo{author}{Dickson, N.},
  \bibinfo{author}{Drew-Brook, M.}, \bibinfo{author}{Rose, G.} \&
  \bibinfo{author}{Aspuru-Guzik, A.}
\newblock \bibinfo{title}{Finding low-energy conformations of lattice protein
  models by quantum annealing}.
\newblock \emph{\bibinfo{journal}{Sci. Rep.}} \textbf{\bibinfo{volume}{2}},
  \bibinfo{pages}{571} (\bibinfo{year}{2012}).

\bibitem{Gaitan2012}
\bibinfo{author}{Gaitan, F.} \& \bibinfo{author}{Clark, L.}
\newblock \bibinfo{title}{Ramsey numbers and adiabatic quantum computing}.
\newblock \emph{\bibinfo{journal}{Phys. Rev. Lett.}}
  \textbf{\bibinfo{volume}{108}}, \bibinfo{pages}{010501}
  (\bibinfo{year}{2012}).

\bibitem{PerdomoOrtiz_EPJST2015}
\bibinfo{author}{{Perdomo-Ortiz, A.}}, \bibinfo{author}{{Fluegemann, J.}},
  \bibinfo{author}{{Narasimhan, S.}}, \bibinfo{author}{{Biswas, R.}} \&
  \bibinfo{author}{{Smelyanskiy, V.N.}}
\newblock \bibinfo{title}{A quantum annealing approach for fault detection and
  diagnosis of graph-based systems}.
\newblock \emph{\bibinfo{journal}{Eur. Phys. J. Special Topics}}
  \textbf{\bibinfo{volume}{224}}, \bibinfo{pages}{131--148}
  (\bibinfo{year}{2015}).

\bibitem{RieffelQIP2015}
\bibinfo{author}{Rieffel, E.~G.} \emph{et~al.}
\newblock \bibinfo{title}{A case study in programming a quantum annealer for
  hard operational planning problems}.
\newblock \emph{\bibinfo{journal}{Quantum Information Processing}}
  \textbf{\bibinfo{volume}{14}}, \bibinfo{pages}{1--36} (\bibinfo{year}{2015}).

\bibitem{OGormanEPJST2015}
\bibinfo{author}{{O'Gorman, B.}}, \bibinfo{author}{{Babbush, R.}},
  \bibinfo{author}{{Perdomo-Ortiz, A.}}, \bibinfo{author}{{Aspuru-Guzik, A.}}
  \& \bibinfo{author}{{Smelyanskiy, V.}}
\newblock \bibinfo{title}{Bayesian network structure learning using quantum
  annealing}.
\newblock \emph{\bibinfo{journal}{Eur. Phys. J. Special Topics}}
  \textbf{\bibinfo{volume}{224}}, \bibinfo{pages}{163--188}
  (\bibinfo{year}{2015}).

\bibitem{Ronnow25072014}
\bibinfo{author}{R{\o}nnow, T.~F.} \emph{et~al.}
\newblock \bibinfo{title}{Defining and detecting quantum speedup}.
\newblock \emph{\bibinfo{journal}{Science}} \textbf{\bibinfo{volume}{345}},
  \bibinfo{pages}{420--424} (\bibinfo{year}{2014}).

\bibitem{Boixo_arXiv2015}
\bibinfo{author}{Boixo, S.} \emph{et~al.}
\newblock \bibinfo{title}{Computational role of multiqubit tunneling in a
  quantum annealer}.
\newblock \emph{\bibinfo{journal}{arXiv:1502.05754}}  (\bibinfo{year}{2015}).

\bibitem{Pudenz_NatCom2014}
\bibinfo{author}{Pudenz, K.~L.}, \bibinfo{author}{Albash, T.} \&
  \bibinfo{author}{Lidar, D.~A.}
\newblock \bibinfo{title}{Error-corrected quantum annealing with hundreds of
  qubits}.
\newblock \emph{\bibinfo{journal}{Nat Commun}} \textbf{\bibinfo{volume}{5}}
  (\bibinfo{year}{2014}).

\bibitem{boixo_NatCommun2013}
\bibinfo{author}{Boixo, S.}, \bibinfo{author}{Albash, T.},
  \bibinfo{author}{Spedalieri, F.~M.}, \bibinfo{author}{Chancellor, N.} \&
  \bibinfo{author}{Lidar, D.~A.}
\newblock \bibinfo{title}{Experimental signature of programmable quantum
  annealing}.
\newblock \emph{\bibinfo{journal}{Nat Commun}} \textbf{\bibinfo{volume}{4}}
  (\bibinfo{year}{2013}).

\bibitem{boixo_evidence_2014}
\bibinfo{author}{Boixo, S.} \emph{et~al.}
\newblock \bibinfo{title}{Evidence for quantum annealing with more than one
  hundred qubits}.
\newblock \emph{\bibinfo{journal}{Nature Physics}}
  \textbf{\bibinfo{volume}{10}}, \bibinfo{pages}{218--224}
  (\bibinfo{year}{2014}).

\bibitem{Shin_arXiv2014}
\bibinfo{author}{Shin, S.~W.}, \bibinfo{author}{Smith, G.},
  \bibinfo{author}{Smolin, J.~A.} \& \bibinfo{author}{Vazirani, U.}
\newblock \bibinfo{title}{Comment on "distinguishing classical and quantum
  models for the d-wave device"}.
\newblock \emph{\bibinfo{journal}{arXiv:1404.6499v2}}  (\bibinfo{year}{2014}).

\bibitem{Albash_EPJST2015}
\bibinfo{author}{{Albash, T.}}, \bibinfo{author}{{R√∏nnow, T.F.}},
  \bibinfo{author}{{Troyer, M.}} \& \bibinfo{author}{{Lidar, D.A.}}
\newblock \bibinfo{title}{Reexamining classical and quantum models for the
  d-wave one processor}.
\newblock \emph{\bibinfo{journal}{Eur. Phys. J. Special Topics}}
  \textbf{\bibinfo{volume}{224}}, \bibinfo{pages}{111--129}
  (\bibinfo{year}{2015}).

\bibitem{MartinMayor_arXiv2015}
\bibinfo{author}{Martin-Mayor, V.} \& \bibinfo{author}{Hen, I.}
\newblock \bibinfo{title}{Unraveling quantum annealers using classical
  hardness}.
\newblock \emph{\bibinfo{journal}{arXiv:1502.02494}}  (\bibinfo{year}{2015}).

\bibitem{Hen_arXiv2015}
\bibinfo{author}{Hen, I.} \emph{et~al.}
\newblock \bibinfo{title}{Probing for quantum speedup in spin glass problems
  with planted solutions}.
\newblock \emph{\bibinfo{journal}{arXiv:1502.01663}}  (\bibinfo{year}{2015}).

\bibitem{King_arXiv2015}
\bibinfo{author}{King, A.~D.}
\newblock \bibinfo{title}{Performance of a quantum annealer on range-limited
  constraint satisfaction problems}.
\newblock \emph{\bibinfo{journal}{arXiv:1502.02098}}  (\bibinfo{year}{2015}).

\bibitem{Katzgraber_PRX2015}
\bibinfo{author}{Katzgraber, H.~G.}, \bibinfo{author}{Hamze, F.} \&
  \bibinfo{author}{Andrist, R.~S.}
\newblock \bibinfo{title}{Glassy chimeras could be blind to quantum speedup:
  Designing better benchmarks for quantum annealing machines}.
\newblock \emph{\bibinfo{journal}{Phys. Rev. X}} \textbf{\bibinfo{volume}{4}},
  \bibinfo{pages}{021008} (\bibinfo{year}{2014}).

\bibitem{Venturelli_arXiv2014}
\bibinfo{author}{Venturelli, D.} \emph{et~al.}
\newblock \bibinfo{title}{Quantum optimization of fully-connected spin
  glasses}.
\newblock \emph{\bibinfo{journal}{arXiv:1406.7553}}  (\bibinfo{year}{2014}).

\bibitem{harris2010}
\bibinfo{author}{Harris, R.} \emph{et~al.}
\newblock \bibinfo{title}{Experimental investigation of an eight-qubit unit
  cell in a superconducting optimization processor}.
\newblock \emph{\bibinfo{journal}{Phys. Rev. B.}}
  \textbf{\bibinfo{volume}{82}}, \bibinfo{pages}{024511}
  (\bibinfo{year}{2010}).

\bibitem{Barahona1982}
\bibinfo{author}{Barahona, F.}
\newblock \bibinfo{title}{On the computational complexity of ising spin glass
  models}.
\newblock \emph{\bibinfo{journal}{J. Phys. A: Math. Gen.}}
  \textbf{\bibinfo{volume}{15}}, \bibinfo{pages}{3241--3253}
  (\bibinfo{year}{1982}).

\bibitem{King_arXiv2014}
\bibinfo{author}{King, A.~D.} \& \bibinfo{author}{McGeoch, C.~C.}
\newblock \bibinfo{title}{Algorithm engineering for a quantum annealing
  platform}.
\newblock \emph{\bibinfo{journal}{arXiv:1410.2628}}  (\bibinfo{year}{2014}).

\bibitem{Babbush2014}
\bibinfo{author}{Babbush, R.}, \bibinfo{author}{Perdomo-Ortiz, A.},
  \bibinfo{author}{O'Gorman, B.}, \bibinfo{author}{Macready, W.} \&
  \bibinfo{author}{Aspuru-Guzik, A.}
\newblock \emph{\bibinfo{title}{Construction of Energy Functions for Lattice
  Heteropolymer Models: Efficient Encodings for Constraint Satisfaction
  Programming and Quantum Annealing}}, \bibinfo{pages}{201--244}
  (\bibinfo{publisher}{John Wiley Sons, Inc.}, \bibinfo{year}{2014}).

\bibitem{Cai-14}
\bibinfo{author}{Cai, J.}, \bibinfo{author}{Macready, B.} \&
  \bibinfo{author}{Roy, A.}
\newblock \bibinfo{title}{A practical heuristic for finding graph minors}.
\newblock \emph{\bibinfo{journal}{arXiv:1406.2741}}  (\bibinfo{year}{2014}).

\bibitem{Choi2008}
\bibinfo{author}{Choi, V.}
\newblock \bibinfo{title}{Minor-embedding in adiabatic quantum computation: I.
  the parameter setting problem}.
\newblock \emph{\bibinfo{journal}{arXiv:0804.4884}}  (\bibinfo{year}{2008}).

\bibitem{PerdomoOrtiz_arXiv2015b}
\bibinfo{author}{Perdomo-Ortiz, A.} \emph{et~al.}
\newblock \bibinfo{title}{Determination of correctable persistent biases in
  quantum annealers}.
\newblock \emph{\bibinfo{journal}{In Preparation}}  (\bibinfo{year}{2015}).

\bibitem{Choi2011}
\bibinfo{author}{Choi, V.}
\newblock \bibinfo{title}{Minor-embedding in adiabatic quantum computation:
  {II}. minor-universal graph design}.
\newblock \emph{\bibinfo{journal}{Quantum Information Processing}}
  \textbf{\bibinfo{volume}{10}}, \bibinfo{pages}{343--353}
  (\bibinfo{year}{2011}).

\bibitem{TrevorJposComment}
\emph{\bibinfo{journal}{Personal communication with T. Lanting. K. Pudenz, and
  S. Adachi.}}  (\bibinfo{year}{2014}).

\end{thebibliography}

\end{document}